\def\sqiglt{\hbox{\rlap{\lower.55ex \hbox{$\sim$}}\kern-.0em \raise.4ex \hbox{$<$}\,}}
\def\sqiggt{\hbox{\rlap{\lower.55ex \hbox{$\sim$}}\kern-.3em \raise.4ex \hbox{$>$}\,}}
\def\sqig{$\sim\,$} \def\etal{et\,al.} \def\msun{M$_{\scriptstyle\odot}$} 
  
 \def\pten#1{$\times10^{#1}$}
\def\kmps{km\,s$^{-1}$} \def\kev{ke\kern-.1em V}

\def\chandra{{\sl Chandra\/}}\def\astroe{{\sl Astro-E2\/}}
\def\asca{{\sl ASCA\/}}\def\ev{e\kern-.1em V}\def\kalpha{{K$\alpha$}}
\documentstyle{mn}
\title[The iron K$\alpha$ complex]
{On the iron K$\alpha$ complex in magnetic cataclysmic variables}
\author[C. Hellier \&\ K. Mukai]
{Coel Hellier$^{1}$ \&\ Koji Mukai$^{2}$\\
$^{1}$School of Chemistry and Physics, Keele University, Staffordshire, 
ST5 5BG\\
$^{2}$Laboratory for High Energy Astrophysics, NASA/GSFC, Code 662, Greenbelt, 
MD 20771, USA}
\begin{document}
\maketitle
\begin{abstract}
We present a compilation of spectra of the iron \kalpha\ region in
magnetic cataclysmic variables, using data from the \chandra -HETG.
The H-like, He-like and fluorescent components are clearly resolved,
and there are hints of the structure within each component. The
different shape of the He-like component in AM Her might be
related to greater cyclotron cooling in this star. A surprising
absence of Doppler shifts in the H-like and He-like components implies
that the X-ray emission is predominantly from the denser, lower-velocity 
base of the accretion columns. This absence will allow \astroe\ to resolve the
structure in each component, leading to temperature diagnostics.

We do not confirm the report that the H-like and He-like components of
AO~Psc are Compton broadened; however we do detect a Compton-downshifted
shoulder to the fluorescent line of GK~Per. Further, a Doppler-shifted
wing of this line arises in the high-velocity, pre-shock flow. 
\end{abstract}
 
\section{Introduction}
The iron \kalpha\ complex is the strongest emission line in the X-ray
spectra of magnetic cataclysmic variables (MCVs). In such stars an
accretion shock above the white dwarf surface heats accreting material
to \sqig 30 \kev. The highly ionized, optically thin plasma then cools
and settles onto the white dwarf, eventually becoming optically thick.
The observed X-ray emission is thus the sum of contributions from a
range of temperatures, densities, and optical depths (e.g., Aizu 1973;
Cropper \etal\ 1999).

In principle, the emission lines will provide diagnostics of the
accretion flow in MCVs [see Fujimoto \&\ Ishida (1997) for a 
determination of white dwarf mass, and Mauche,
Liedahl \&\ Fournier (2001; 2003) for density diagnostics] though the
inhomogeneity of the accretion column complicates the interpretation.
The relative contributions of collisional ionization and
photo-ionization are also uncertain: Mukai \etal\ (2003) argue that
photo-ionization dominates the spectrum at high mass-accretion rates
but not at lower rates.

Early studies of the Fe \kalpha\ line in MCVs (e.g.\ Norton, Watson
\&\ King 1991) used data that had insufficient resolution to resolve
the components from H-like, He-like and colder iron.  This
was first possible with the CCD detectors on the \asca\ satellite,
which gave 120-\ev\ resolution. Using \asca\ data, Hellier, Mukai \&\
Osborne (1998) claimed that in some MCVs (e.g.\ V1223~Sgr) the three
components were clearly resolved, whereas in others (e.g.\ AO~Psc) the
components appeared broadened and blended. The \sqig 150-\ev\ 
broadening was attributed to Compton scattering.

The high-energy transmission grating (HETG) on \chandra\ is a further
advance: though it has a lower effective area than the \asca\ CCDs,
its 35-\ev\ resolution is sufficient to separate the three main
components with ease, and to discern some of the structure within the
components.  The forthcoming \astroe\ will be better still, with
6-\ev\ resolution and a high effective area.  In this paper we report
the appearance of the Fe \kalpha\ complex at the \chandra-HETG resolution,
based on a compilation of MCV spectra.

\section{Observations and results}
Five MCVs have so far been observed with the \chandra -HETG, namely
AO~Psc, AM~Her, EX~Hya, V1223~Sgr and GK~Per.  The observations are
listed in Table~1. For each we extracted the HEG (high energy) and MEG
(medium energy) spectra, combining the +1 and --1 orders.

The GK~Per data consist of two observations taken during the 2002
outburst when the optical magnitudes were 11.0 and 10.5. In our
analysis we have summed the two spectra.  The AM~Her spectra were
obtained in an intermediate brightness state at an optical magnitude
of 14; the AO~Psc, V1223~Sgr and EX~Hya spectra were obtained in the
usual state of these stars.

\begin{figure*}\vspace*{19cm}     
\caption{The iron \kalpha\ region of \chandra\ HEG spectra of 5 MCVs.
At top we show the locations of the 
Ly$\alpha_{1}$ and Ly$\alpha_{2}$ components of H-like iron; the 
resonance, intercombination and forbidden components of He-like iron;
and the fluorescence line of cold iron. The dashed lines show the
locations of major dielectronic satellite lines. The Gaussian illustrates the
nominal resolution of the HEG at this energy. All spectra are binned
to a minimum of 30 counts per bin (50 for GK Per), and thus  
have a typical signal-to-noise ratio (S/N) of 6 (8 for GK Per). A 
typical error bar is shown near 6 \kev.}
\includegraphics{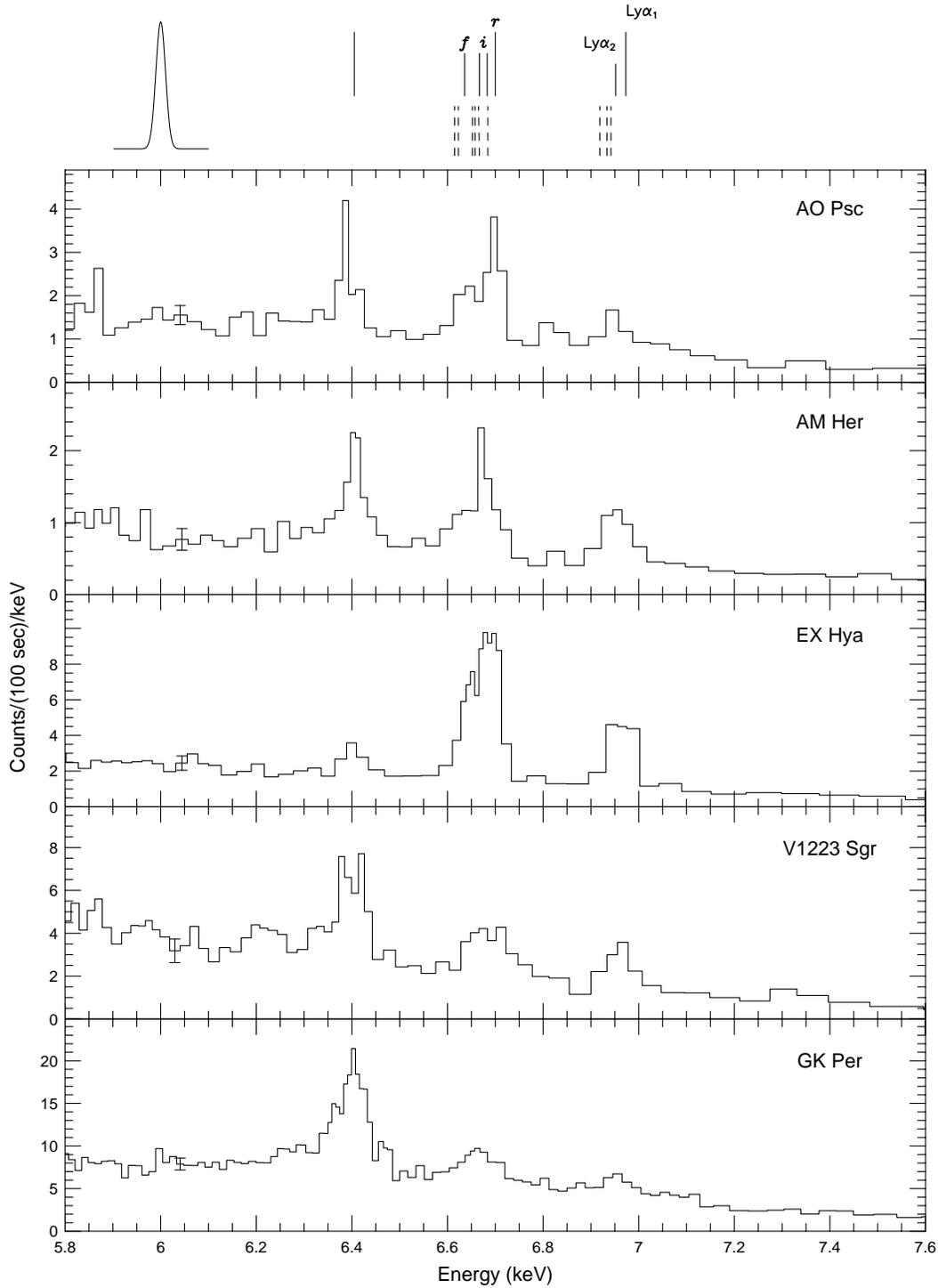} 
\end{figure*}

\begin{table}
\begin{tabular}{lc@{}c@{\hspace{5mm}}r@{\hspace{7mm}}r@{\hspace{7mm}}rr}
\hline
Star & Date & Exp.~(ks) & \multicolumn{4}{c}{\hspace{-5mm}Equivalent widths (\ev)}  \\ 
  &  &  & \multicolumn{4}{l}{\hspace{-5mm}6.4-\kev\ \ 6.7-\kev\ \ 6.97-\kev} \\ \hline
AO Psc & 2001/05/23 & 98 & 101 & 165 & 138 \\ [1mm]
AM Her & 2003/08/15 & 93 &  120 & 148 & 119 \\ [1mm]
EX Hya & 2000/05/18 & 59 & 35  & 322 & 130 \\ [1mm]
V1223 Sgr & 2000/04/30 & 51 &  108 & 67  & 87  \\ [1mm]
GK Per &  2002/03/27 & 32 &  260 & 117 & 80 \\ 
       &  2002/04/09 & 34 \\ \hline
\end{tabular}
\caption{The observations and measured equivalent widths.}
\end{table}

Fig.~1 shows the iron \kalpha\ region of HEG data for each star (the
MEG data are not useful at this energy and so are not shown here, but
see Mukai \etal\ 2003).

\begin{figure}\vspace*{10cm}     
\caption{The iron \kalpha\ region of AO Psc as observed with the \asca\
SIS (top) and the \chandra\ HEG (bottom). The illustrative model is a
power law and three Gaussians (at 6.38, 6.67 and 6.96 \kev, all of
width 35 eV).  The fit parameters are optimised for the \chandra\ data alone.}
\includegraphics{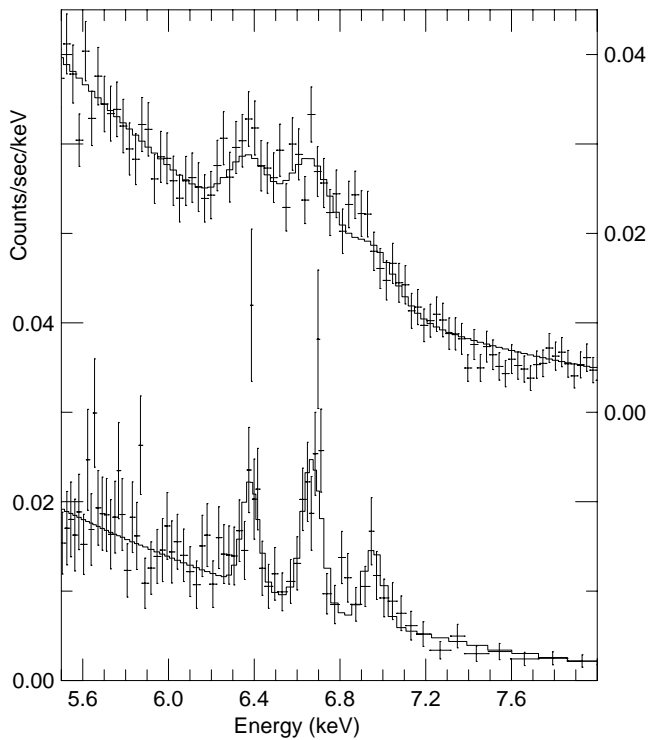} 
\end{figure}

The division into three distinct components is clear in all cases,
conflicting with Hellier \etal's (1998) report of blended lines,
in particular in the \asca\ SIS data of AO Psc.  To investigate this further 
we show in Fig.~2 both the \chandra\ HEG and the \asca\ SIS spectra of
AO Psc. The illustrative model is a power law and three Gaussians 
fitted to the \chandra\ data. The same model (adjusting only the 
power-law normalisation) is shown plotted on the \asca\ data after
folding through the \asca\ response. 

The \asca\ spectrum has a group of 4 high points between the 
H-like and He-like lines, explaining why the \asca\ spectrum is
best fit by very broad, blended H-like and He-like lines. This
is not supported by the higher-resolution \chandra\ data (though
there are two high points in a similar place).  Overall we 
suggest that apparent broadening in the \asca\ spectrum is a 
statistical fluke, although a change in the source characteristics 
cannot be ruled out, and we await an \astroe\ spectrum to finally
settle the issue. 

It is also possible that the narrow lines seen in the \chandra\ data
are accompanied by a broad, faint, Compton-scattered base that is hard
to detect in high-resolution but low-S/N grating spectra.  Indeed,
this is expected to some degree (see Hellier \etal\ 1998), since the
resonance lines are likely to be borderline optically thick,
increasing the path length of resonance-line photons to the point
where Compton scattering is appreciable (see also Terada \etal\ 2001 for 
the effect of resonance scattering on equivalent widths).

\section{The H-like line}
The `6.97-\kev' line of Fe {\sc xxvi} consists of Ly$\alpha_{1}$ at
6.973 \kev\ and, with theoretically half the intensity, Ly$\alpha_{2}$
at 6.952 \kev\ [these are transitions to $1s(^{2}\!S_{1/2})$ from
$2p(^{2}\!P_{3/2})$ and $2p(^{2}\!P_{1/2})$ respectively].  Dielectronic
recombination leads to satellite lines at slightly lower energies in
the range 6.91--6.95 \kev\ (the
energy shift arises from the presence of a `spectator' electron in a
higher-$n$ level; strictly they are He-like lines, but occur near the
H-like line since the second electron is not in the ground state and
so doesn't perturb the energy as much).  Calculations of the positions
and intensities of the satellite lines are given by Dubau
\etal\ (1981) while observations of the H-like complex in the solar
spectrum can be found in Pike \etal\ (1996).

We can thus ask whether the satellite lines are contributing to the
MCV spectra, Although we can't resolve them at the HEG resolution,
their presence would shift the line centroid to the red.
Thus we used XSPEC's {\sc fakeit} facility to find that narrow  
Ly$\alpha_{1}$ and Ly$\alpha_{2}$ components alone would, folded
through the HEG response and fitted with a Gaussian, produce a line 
centroid of 6.964 \kev\ with a width ($\sigma$) of 9 \ev. 

Fitting the observed 6.97-\kev\ lines with a Gaussian produced a redder
centroid. A fit to all 5 stars simultaneously found a centroid of
6.958\,$\pm$\,0.003 \kev, with 6.964 ruled out at 90 per cent
confidence.  The measured width, at 24\,$\pm$\,6 \ev, was
significantly greater than 9 \ev.

To interpret this we need to consider possible Doppler shifts, since
the 6-\ev\ shift is equivalent to 260 \kmps\ and the broadening to 1000 
\kmps. We thus repeated the analysis on the H-like lines of Ne, Mg and
Si, using the MEG data. Since the strength of satellite lines scales
as $z^{4}$, these lines should be free of satellites.  We found no
shifts to a 90-per-cent limit of 100 \kmps\ and no broadening to a
limit of 450 \kmps. This doesn't completely exclude Doppler shifts in
the Fe lines, since the Fe emission centroid will be hotter and 
higher up the accretion column, where velocities are larger. However,
the low level of any Doppler shifts suggests that the observed shift
in H-like Fe is instead caused by the presence of satellite lines.

The lack of Doppler shifts to a limit of 100 \kmps\ is somewhat
surprising, considering that the escape velocity of a 0.7-\msun\ white
dwarf is \sqig 5000 \kmps. However, the infall velocity drops by a
factor 4 through the stand-off shock, and reduces further as the
material cools and settles onto the white dwarf (scaling with height
as $h^{2/5}$; e.g., Frank, King \&\ Raine 2002).  Since the density
scales inversely to the velocity, and since the X-ray emissivity
depends on density squared, the emission is predominantly from
low-velocity material close to the white dwarf surface. Further, the
projection onto the line of sight, and the fact that the projected
motions will be smeared out over the spin cycle, can reduce the
observed velocities by another factor 2--3.  Thus the observed
100-\kmps\ limit is explained if the emission arises predominantly
from the lowest few per cent of the accretion column.

\section{The He-like line}
The `6.7-\kev' line of Fe {\sc xxv} consists of the resonance line,
$r$, a forbidden line, $f$, and two intercombination lines, $i$ [being
transitions to the ground state $1s^{2}(^{1}\!S_{0})$ from
$1s2p(^{1}\!P_{1})$, $1s2s(^{3}\!S_{1})$ and $1s2p(^{3}\!P_{2,1})$
respectively].  Interspersed with these are dielectronic satellite
lines over the range 6.61--6.68 \kev.  The relative line intensities
depend on temperature, density, the degree of photoionization, and the
extent to which the resonance line is depleted by Compton scattering
(which is greatly enhanced by resonant trapping of such photons).
Calculations of the 6.7-\kev\ complex can be found in Bautista \&\
Kallman (2000) and Oelgoetz \&\ Pradhan (2001). Such calculations
indicate that satellite lines will dominate the 6.7-\kev\ complex at
temperatures below 3\pten{7}\,K, with the He-triplet components
dominating at higher temperatures.

At the \chandra\ resolution the different components are not resolved,
but some tentative statements can be made.  The AO~Psc spectrum is
compatible with a strong resonance component and contributions from
the forbidden lines, or nearby satellites.  The relative strength of
the resonance line implies a temperature of $>$\,3\pten{7}\,K [see the
diagnostic curves in Bautista \&\ Kallman (2000) and Oelgoetz \&\
Pradhan (2001)].

In contrast, in the AM~Her spectrum the resonance line appears weaker,
with the intercombination lines, or adjacent satellites, appearing
strongest. One possibility is that resonant scattering is affecting
this line (a resulting beaming of resonance-line photons is
already thought to be important in some AM Her stars; Terada \etal\ 2001).

Alternatively, a weak resonance line could result from a low
temperature of $<$\,2\pten{7}\,K ($<$\,2\,\kev).  At such temperatures
the line will be dominated by satellite lines, assuming that
collisional equilibrium holds. A low temperature is consistent with
the lack of Doppler shifts, since $v/v_{\rm shock} = T/T_{\rm shock}$,
so that the observed 100-\kmps\ limit implies a temperature in the
line emitting region of \sqiglt 0.2\,$T_{\rm shock}$.  On the other
hand, Ishida \etal\ (1997) deduce a temperature of 10--15 \kev\ from
the line ratios observed in the \asca\ spectrum of AM~Her, which is
incompatible with our interpretation.  Both of these temperature
estimates are based on barely-resolved line data, so we await \astroe\
for more reliable results.

A possible explanation for AM Her being different from AO Psc is that
AM Her, being a polar, has a much stronger magnetic field of 14 MG
(Bailey, Ferrario \&\ Wickramasinghe 1991), whereas the other stars
have weaker fields of \sqig 1 MG.  Thus cyclotron cooling will be
important in AM Her, but not in the others.

The EX~Hya data suggest a strong resonance line, along with $i$, and $f$
and/or satellites. Overall, the ratio of He-like/H-like equivalent
widths is very different from that in the other systems
(Table~1). Similarly, Mukai \etal\ (2003) found that the EX~Hya
spectrum is unlike that of other magnetic MCVs, being
compatible with a collisionally-excited, cooling-flow model, whereas
others showed photoionized spectra. The difference is probably due to
the much lower accretion rate and luminosity of EX~Hya, which are in
line with it being the only system in our sample below the period
gap.  

\section{The fluorescence line}
The fluorescence line peaks near 6.41 \kev\ in all stars, compatible
with cold iron in states Fe {\sc i} to Fe {\sc xvii}.  There is little
sign of iron at ionizations between {\sc xvii} and {\sc xxv}.

The GK~Per fluorescent line has the highest S/N. It has a red wing
extending to 6.33 \kev, consistent with a Doppler shift of up to 3700
\kmps.  We suggest that this arises from pre-shock material, which
will be falling at near the white-dwarf escape velocity.

Further, the GK~Per line appears to have a fainter shoulder to its red
wing, extending to 170 \ev\ from the line centre.  We suggest that
these are photons Compton downscattered by up to two Compton wavelengths.

It is notable that GK~Per's fluorescent line has an equivalent width
more than a factor two greater than seen in the other systems (Table~1).
Since this line arises from scattering of the emitted X-rays, the 
equivalent width will be set primarily by the size of the scattering
targets, namely the white dwarf (filling half the solid angle as seen
from the emission region) and the inflowing material.  

In most MCVs material accretes only over a small range of magnetic
longitude, as deduced from eclipse studies of the size of the emission
region (e.g.\ Hellier 1997), and from the blackbody areas of regions
of heated white-dwarf surface (e.g.\ Haberl \&\ Motch 1995).  Thus the
`accretion curtains' of infalling material are a small target compared
to the white dwarf, in keeping with the fact that the fluorescent lines
are not Doppler-shifted wholesale.

However, from an analysis of the spin-cycle pulsation of GK~Per,
Hellier, Harmer \&\ Beardmore (2004) claimed that, during outburst,
the material accretes from all azimuths to form a complete accretion
ring. This would fill most of the sky, as seen from the emission
region, thus greatly enhancing the equivalent width of the fluorescent
line. In keeping with this idea the fluorescent line observed in
quiescence by \asca\ is much weaker, with an equivalent width of only
50 \ev\ (Hellier \etal\ 1998).

\section{Conclusions}
We present the first compilation of Fe \kalpha -line spectra
of MCVs using the \chandra -HETG. We conclude that:

(1) Dielectronic satellite lines are detected in the H-like and
He-like iron \kalpha\ lines. These cause the lines to be broadened
and redshifted, as predicted by Oelgoetz \&\ Pradhan (2001).

(2) The H-like lines show no Doppler shifts to a limit of 100 \kmps.
This implies that the emission is predominantly from the lowest few
per cent of the accretion columns, where the density is highest.

(3) The H-like and He-like lines of AO~Psc are not Compton broadened
by 150 \ev, as had been suggested from a lower-resolution \asca\ spectrum.

(4) The He-like lines in AO~Psc and AM~Her show a discernably
different structure. This could be caused by greater
cyclotron cooling in AM~Her. 

(5) The absence of Doppler shifts and the hints of structure in the
He-like line are encouraging for observations with the XRS
on the forthcoming \astroe.  Its 6-\ev\ resolution will 
separate the different He-triplet and satellite lines, allowing 
the application of temperature diagnostics. 

(6) The fluorescence line in GK~Per shows a Doppler-shifted red wing
owing to pre-shock material falling at near the escape velocity.  We
also detect a fainter, 170-\ev\ shoulder caused by Compton
down-scattering.  The high equivalent width of GK~Per's fluorescence
line supports the suggestion by Hellier \etal\ (2004) that, during
outburst, the accretion flows from all azimuths to form a complete
accretion ring.

\end{document}